\documentclass[a4paper,11pt]{article}
\usepackage{multirow}
\usepackage{jcappub} % for details on the use of the package, please see the JINST-author-manual
\usepackage{lineno}
\newcommand{\be}{\begin{equation}}
\newcommand{\ee}{\end{equation}}
\newcommand{\ba}{\begin{eqnarray}}
\newcommand{\ea}{\end{eqnarray}}
\newcommand{\bi}{\begin{itemize}}
\newcommand{\ei}{\end{itemize}}

\title{On the capability of high redshift kSZ measurement with galaxy surveys}
\author[a,b]{Ziyang Chen}
\author[a,c,b]{Pengjie Zhang}

\affiliation[a]{Department of Astronomy, School of Physics and Astronomy, Shanghai Jiao Tong University, Shanghai, China}
\affiliation[b]{Key Laboratory for Particle Astrophysics and Cosmology (MOE)/Shanghai Key Laboratory for Particle Physics and Cosmology, 
Shanghai, China}
\affiliation[c]{Tsung-Dao Lee Institute, Shanghai Jiao Tong University , Shanghai
200240, China}

\emailAdd{chen\_zy@sjtu.edu.cn; zhangpj@sjtu.edu.cn}

\abstract{The kinematic Sunyaev-Zel’dovich (kSZ) effect has been detected at $z<1$ using various techniques and data sets. The ongoing and upcoming spectroscopic galaxy surveys such as DESI (Dark Energy Spectroscopic Instrument) and PFS (Prime Focus Spectrograph) will push the detection  beyond $z = 1$, and therefore map the baryon distribution at high redshifts. Such detection can be achieved by both the kSZ stacking and tomography methods. While the two methods are theoretically equivalent, they differ significantly in the probed physics and scales, and required data sets. Taking the combination of PFS and ACT (Atacama Cosmology Telescope) as an example, we build mocks of kSZ and galaxies, quantify the kSZ detection S/N,  and compare between the two methods. 
We segment the PFS galaxies into three redshift bins: $0.6 < z < 1.0$, $1.0 < z < 1.6$, and $1.6 < z < 2.4$. For tomography method, our analysis reveals that the two higher redshift bins exhibit significantly higher S/N ratios, with values of 32 and 28, respectively, compared to the first redshift bin, which yielded an S/N of 8. This is attributed to not only the increasing of electron density with redshifts, but also the larger survey volume and the reduced non-linearity, facilitating velocity reconstruction at higher redshifts. Therefore, the capability of the PFS survey to measure high redshift kSZ effect stands as a substantial advantage over other spectroscopic surveys at lower redshift. The S/N of kSZ stacking largely depends on the number of galaxy groups available from another photometric survey. But in general, its S/N is lower than that of kSZ tomography, largely due to CMB instrument noise and error in galaxy group redshift. 
Incorporating next-generation CMB surveys like CMB-S4, characterized by significantly reduced instrument noise and improved angular resolution, is expected to enhance tomographic detection by a factor of ten and stacking detection by fivefold. This future high S/N detection holds the promise of not only providing precise constraints on the overall baryon abundance but also initiating a new insight into baryon distribution.

}


\begin{document}
\maketitle
\flushbottom

\section{Introduction} \label{sec:introduction}

    In cosmology, the kinematic Sunyaev Zel'dovich (kSZ) effect stands as a pivotal probe for understanding the large-scale baryon distribution of the universe.  The kSZ effect, a secondary anisotropy in the CMB, arises from the inverse Compton scattering of CMB photons off a group of free electrons moving with respect to the CMB rest frame. Unlike the thermal Sunyaev-Zel'dovich (tSZ) effect and X-ray emissions, the kSZ effect linearly depends on electron density and is independent of electron temperature, making it a valuable probe for mapping baryon in less dense regions of the universe, such as filaments and voids. This probe has a powerful constraining ability in colder and the context of "missing baryon" problem in the local universe, where observations \cite{Fukugita2004} suggest a $50\%$ deficit relative to the predictions of Big Bang Nucleosynthesis (BBN) \cite{Bregman2007}.  On the other hand, if the baryon distribution can be constrained by other observations, such as tSZ \cite{battaglia2016}, Fast Radio Bursts \cite{madhavacheril2019a}, the kSZ effect encodes valuable information about the growth rate \cite{planckcollaboration2014a, alonso2016a} of the universe's matter content, leveraging the investigation of the dark energy \cite{mueller2015}, modified gravity \cite{bianchini2016a, mitchell2021a, okumura2022a} and neutrino mass \cite{mueller2015a}.

    Despite its significance, the kSZ signal is notably faint compared to other CMB components, presenting a formidable challenge for accurate detection and measurement. A critical aspect of extracting the kSZ signal hinges on its unique dependence on the peculiar velocity field.
    The first measurement of the kSZ effect by ref.~\cite{hand2012a} utilized the pairwise method, which capitalizes on the gravitational interaction between cluster pairs moving towards each other. While this method provides insights into the relative motion between clusters, it overlooks the collective motion that could yield additional information. Refs.~\cite{planckcollaboration2016, debernardis2017a, li2018b, calafut2021a} further developed this method and gain $1.8\sim 5.4 \sigma$ detection. 
    Ref. ~\cite{Gong2024} constructed the pairwise velocity from the kSZ signal using machine learning, which can break the degeneracy between pairwise velocity and optical depth, and further constrain dark energy and modified gravity theories.
    The advantage of the pairwise method is that it does not require precise knowledge of the clusters' peculiar velocities. 
    In addition, apart from the spectroscopic galaxies, the pairwise method can also utilize enormous number of photo-z galaxies/groups \cite{soergel2016a, chen2022, schiappucci2023a, li2024}, although the large redshift uncertainty causes a large suppression of pairwise kSZ signal especially when $r<100\ {\rm Mpc}/h$ \citep{flender2016a}.
    
    Advancing the measurement of the kSZ effect involves reconstructing peculiar velocities of clusters, from spectroscopic redshift galaxy surveys \cite{li2014a, schaan2016, schaan2021a, mallaby-kay2023a}.  By stacking the CMB signals at cluster locations, the signal-to-noise ratio (S/N) is enhanced. Weighting them by the line-of-sight (LOS) velocities would effectively isolate the kSZ signal from other unrelated CMB components. This approach is particularly effective in studying baryon distributions around clusters. Refs.~\cite{guachalla2023a, hadzhiyska2023a} investigated the effect of various factors on reconstruction efficiency, such as RSD, photometric redshift errors, satellite galaxy fraction, the adoption of incorrect cosmological models, and the scale of smoothing. In addition, they applied this method to a realistic mock, include light cone effect, survey masks and selection functions. In addition, ref.~\cite{tanimura2022a} attempted to reconstruct velocity by Convolutional Neural Network and found performance is slightly improved compared to the conventional method which uses the continuity equation \cite{tanimura2021a}. In addition, Ref.~\cite{nguyen2020a} found the uncertainty in the velocity reconstruction step, which is often neglected, increases the final uncertainty by about 15\% at cluster scale.
      
    The above kSZ stacking method requires pre-existing cluster/group catalogs, other than the spectroscopic redshift galaxy catalogs. Alternatively, the kSZ tomography method, as proposed by ref.~\cite{ho2009a,shao2011a}, requires no such cluster/group catalog. It reconstructs a kSZ template, which integrates the reconstructed momentum field $\mathbf{\hat p} = (1+\hat\delta)\mathbf{\hat v}$ along the LOS with a weight dependent on redshift as an estimator of the true kSZ signal. This estimator, uncorrelated with other CMB components, extracts the kSZ signal by correlating with a CMB map. This technique is adept at assessing the kSZ effect on the large scale, sensitive to baryon distributions in not only clusters/groups, but also filaments and even voids. 

    Moreover, Ref.~\cite{chaves-montero2021} proposed a novel method called AFR-kSZ tomography. Angular Redshift Fluctuation (ARF), first proposed by \cite{HM2021}, is the statistic of sky maps built by projecting redshifts in a thin shell, sensitive to both underlying density and velocity fields. Ref.~\cite{chaves-montero2021} resorted to the 6dF and SDSS, and foreground-cleaned CMB from Planck to conduct a baryon census in a wide redshift range $0 < z < 5$. They found AFR-kSZ tomography measurement is sensitive to more than half of all baryon in the Universe, with a joint detection $> 10 \sigma$.
    
    Existing kSZ measurements have provided useful information on the baryon distribution and evolution in our universe. Ref.~\cite{planckcollaboration2016} sought to quantify the baryon abundance around central galaxies at redshift $z\sim 0.1$ by estimating the pairwise signal using the combination of Planck data and the Central Galaxy Catalogue samples from the SDSS-DR7. Further refining this measurement, ref.~\cite{hernandez-monteagudo2015a} constrained the baryon fraction by increasing the aperture filter radius and reported an increment in the baryon fraction to 45-55\% of the total baryonic content when the filter radius equals to 10 arcmin, albeit with an increasing noise at larger radii. Ref.~\cite{hill2016a} extended the detection of baryon fraction to higher redshift (z = 0.4), cross-correlating between a projected density field and the squared CMB temperature map. This work achieved a $4.5\sigma $ confirmation that the baryon fraction aligns with BBN prediction. The study continued by ref.~\cite{kusiak2021a}, which applied a similar methodology across three unWISE galaxy catalog samples peaking at redshifts of 0.6, 1.1, and 1.5. Their results indicated the free gas fraction for the higher redshift bins is about twice as high than the fiducial value. They attributed the abnormally high kSZ signal amplitude to uncertainties in theoretical modeling. In addition, by measuring the pairwise kSZ signal of clusters in redshift span $0.1 \sim 0.9$, ref.~\cite{soergel2016a} constrained the gas fraction $f_{\rm gas} = 0.086\pm 0.027$ inside the $R_{500}$. Measuring the pairwise kSZ in Fourier space, ref.~\cite{sugiyama2018a} constrained the gas fraction of LOWZ catalog from BOSS DR12, $f_{\rm gas} = 0.17$, while this result has large error contour and has a degeneracy with the assumption of gas profiles.
    Ref.~\cite{lim2020b} extracted the kSZ signal of a group catalog from SDSS DR7 \cite{Yang2007} and Planck. They divided the catalog into six mass bins and measured the kSZ flux inside $R_{200}$ as a function of halo mass. They found the relationship of kSZ flux and halo mass is self-similar, inferring a mass-independent baryon fraction. They also found the baryon fraction of all mass bin is consistent with the universal value within the errorbars and indicates all “missing baryons” are found on the halo scales. The discussion of “missing baryon" problem has been last twenty years and hasn't been settled due to current low measurement significance. Refs.~\cite{schaan2021a, amodeo2021, mallaby-kay2023a} concentrated more on the baryon distribution inside the groups, thanks to the high resolution of ACT survey. They measured the gas density profile by stacking the kSZ signal at the position of a photometric cluster catalog with varying filter sizes to extract kSZ signal from CMB maps, weighting with the line-of-sight (LOS) velocity reconstructed by another spectroscopic galaxy catalog. They found the gas profile is more extended than that of dark matter and pointed out the potential of kSZ effect to measure the feedback in the galaxy formation model \cite{zheng2024}. Recently, Ref.~\cite{Hadzhiyska2024} measured the kSZ profile around clusters using the ACT and DESI photometric galaxies. 
    They found the gas is much more extended than the dark matter at 40 $\sigma$. More interestingly, their results strongly favor high-feedback models in original Illustris instead of low-feedback models in illustrisTNG simulation.
 
    Detection of kSZ through galaxy-CMB cross-correlation so far is limited to $z<1$. Ongoing surveys such as DESI and Euclid and upcoming surveys such as PFS will push the measurement to $z>1$ and reveal the baryons at high redshifts. The capability of kSZ measurement at high redshift is promising due to higher electron number density and potentially better velocity reconstruction. 
    In this study, we forecast the measurement significance of kSZ using the PFS and the CMB surveys such as ACT and CMB-S4. Because FPS provides accurate positions of galaxies with spectroscopic redshifts, which can facilitate the velocity reconstruction, we focus on stacking and tomography methods. These two methods offer complementary insights, measuring the one-halo and two-halo terms of the kSZ signal, respectively. The PFS extends to redshifts of $z=2.4$, enabling high-redshift measurements. Such high-redshift observations can constrain the abundance and evolution of baryons, providing valuable insight in the ‘missing baryon’ problem. Moreover, one-halo term measurements of the kSZ effect could reveal the baryon fraction in clusters as a function of radius, which is sensitive to baryonic processes such as supernova and Active Galactic Nuclei (AGN) feedback, aiding in calibrating hydrodynamic simulations. As ref.~\cite{Chen2023bPe} demonstrates, simulations with varying galaxy formation models align at lower redshifts but diverge significantly beyond $z=2$. Thus, high-redshift kSZ measurements offer invaluable information for understanding the galaxy formation model. 
    
    This paper is organized as follows. In section \ref{sec:vel_rec_method}, we present the velocity reconstruction algorithm used in both stacking and tomography methods. In section \ref{sec:simulation} we introduce the N-body simulation CosmicGrowth and the methods to construct the mock galaxy and galaxy group catalogs and kSZ/CMB maps. Then we show the forecast of the kSZ measurement significance with two methods in section \ref{sec:forecast}. In the end, we conclude in section \ref{sec:conclusion}.

\section{Velocity reconstruction method}\label{sec:vel_rec_method}
    To extract the kSZ signal from other CMB components, the key is to leverage its correlation with the peculiar velocity of the electron bulk, which is absent in other components. 
    Both kSZ stacking and tomography measurement methods involve reconstructing the peculiar velocity field. In this section, we describe the method of velocity reconstruction from a spectroscopic redshift galaxy catalog.
    
    In the linear regime, the large-scale velocity field is determined by the density field through the continuity equation
    \begin{eqnarray}\label{eq:continuity_equation}
        \dot{\delta}_m + \nabla \cdot \mathbf{v} = 0,
    \end{eqnarray}
    where $\delta_m$ is the matter overdensity and $\textbf{v}$ is the peculiar velocity. Furthermore, the time derivative of the matter overdensity is expressed as 
    \begin{eqnarray}
        \dot{\delta}_m = \frac{d\delta_m}{da}\dot{a} = Hf\delta_m,
    \end{eqnarray}
    where $f = d\ln D/d\ln a$ is the growth rate and $H$ is the Hubble parameter. In Fourier space, eq.~ \ref{eq:continuity_equation} transforms to
    \begin{eqnarray}
        \hat{\mathbf{v}}(\mathbf{k}) = -ifH\delta_m(\mathbf{k})\frac{\mathbf{k}}{k^2},
    \end{eqnarray}
    where $k = |\mathbf{k}|$ is the length of the wavenumber.
    In reality, however, the overdensity of dark matter is unknowable. The galaxy overdensity is used as a biased tracer of dark matter, $\delta_g = b_g \delta_m$. Unfortunately, the RSD and shot noise in the galaxy density field would decrease the velocity reconstruction performance. In the following, we use Wiener filters to suppress the noise and optimize the velocity reconstruction.  

    Our algorithm for reconstructing the velocity field proceeds as follows:\\
    \begin{enumerate}
        \item Divide the entire simulation volume into $2048^3$ voxels, and estimate the overdensity field $\delta^{\rm RSD}_g(\mathbf{x})$ from the galaxy catalog with the Nearest Grid Point (NGP) method.\\
        \item Estimate the divergence of the velocity $\theta(\mathbf{x}) \equiv \nabla \cdot \mathbf{v}(\mathbf{x})$ from the density field  $\delta^{\rm RSD}_g(\mathbf{x})$ with a Wiener filter defined as
        \begin{eqnarray}
            W_1(k_\perp, k_\parallel) = 
            \frac{P_{\theta^t, \delta^{\rm RSD}_g}(k_\perp, k_\parallel)}{P_{\delta^{\rm RSD}_g, \delta^{\rm RSD}_g}(k_\perp, k_\parallel)}.
        \end{eqnarray}
        Here, $\theta^t$ represents the true convergence component from the dark matter particles in simulation without RSD, $k_\perp$ is the length of the wavenumber perpendicular to the LOS direction, and $k_\parallel$ the length is the parallel component.
        This Wiener filter $W_1(k_\perp, k_\parallel)$ is a function dependent on the lengths of $k_\perp$ and $k_\parallel$. This filter could be derived from simulations incorporating appropriate cosmological and galaxy formation models.
        And reconstructed convergence component is
        \begin{eqnarray}
            \hat{\theta}(k_\perp, k_\parallel) = H(z)f(z)W_1(k_\perp, k_\parallel)\delta^{\rm RSD}_g, 
        \end{eqnarray}
        where $H(z)$, $f(z)$ is the Hubble parameter and growth factor at the effective redshift of the galaxies.
        \item Estimate the reconstructed velocity field from the reconstructed convergence component $\hat \theta(k_\perp, k_\parallel)$
        \begin{eqnarray}
            \hat{\mathbf v}(k_\perp, k_\parallel) = \frac{-i\mathbf{k}}{|\mathbf{k}|^2}\hat{\theta}(k_\perp, k_\parallel)
        \end{eqnarray}
        And transform it into the Cartesian coordinates $\hat{\mathbf{v}}(\mathbf{x})$.
    \end{enumerate}

    In the stacking method, we obtain the LOS velocity of each galaxy group $\hat v_{{\rm los}, i}$, by interpolating the reconstructed field $\hat{\mathbf{v}}(\mathbf{x})$ at their position $\mathbf{x}^{\rm ph}_{i, {\rm group}}$.

    In the tomography method, the reconstructed momentum field is the product of the reconstructed density and the reconstructed velocity field 
    \begin{eqnarray} \label{eq:electron_momentum}
        \hat{\mathbf{p}}(\mathbf{x}) = (1+\hat\delta(\mathbf{x}))\hat{\mathbf{v}}(\mathbf{x}),
    \end{eqnarray}
    where  $\hat\delta(\mathbf{k}) = W_2(k)\delta^{\rm RSD}_g$. $W_2(k)$ is the Wiener filter defined as 
    \begin{eqnarray}
        W_2(k) = \frac{b^2_g P_{\rm mm}(k)}{b^2_g P_{\rm mm}(k)+\frac{1}{\bar n_g}},
    \end{eqnarray}
    where $P_{\rm mm}(k)$ is the matter power spectrum and $\bar n_g$ is the mean galaxy number density.\\

    The two Wiener filters, $W_1(k_\perp, k_\parallel)$ and $W_2(k)$, used to mitigate the noise from RSD and shot noise, have been demonstrated theoretically to be the optimal choice in appendix~\ref{appkSZMethodEquivalence}. And comparing to conventional Gaussian filter, the Wiener filters can improve the tomography method when $k>0.1 h/{\rm Mpc}$ (appendix~\ref{app:test_vel_rec}).
    
\section{Mock generation} \label{sec:simulation}
    The PFS is a cosmology survey that offers unique capabilities in astronomical observations due to its large field of view and impressive multiplexing capability, which enables high sampling rates and coverage of a large volume of space \cite{Takada2014}.  It is designed to explore a previously uncharted epoch in cosmic history, extending to redshift $z=2.4$. The sky coverage is $1400\ {\rm deg^2}$. Notably, the PFS survey can cover a comoving volume that is ten times larger than SDSS and twice as large as BOSS, which allows for more detailed studies of cosmic structures and the dark energy component of the universe. The PFS targets a range of galaxy types using optical and near-infrared wavelengths. Specifically, [O II] emission-line galaxies (ELGs), which are useful tracers for conducting an efficient survey out to high redshifts beyond z=1.
    
    In this study, we investigate the potential for detecting the kSZ effect with the PFS ELG catalog.
    We divide the whole redshift range covered by PFS into three bins: $0.6 < z < 1.0$, $1.0 < z < 1.6$ and $1.6 < z < 2.4$. The comoving intervals of these three bins are almost equivalent. The parameters for the PFS cosmology survey for these three bins are listed in table~\ref{Table:PFS_parameter}. 
    In this section, we describe the generation of mock kSZ maps, PFS galaxy catalogs and galaxy group catalogs from the CosmicGrowth simulation used to forecast the kSZ measurement significance.
    
    \begin{table*}
    \centering
    \begin{tabular}{cccccccccc}
    \hline
    & Redshift bin & $\bar n_g \times 10^{-4} [{\rm Mpc}/h]^{-3}$ & $V_{\rm survey} [{\rm Gpc}/h]^3$ &  Snapshot & HOD model\\
    \hline
    1 & $0.6 < z < 1.0$ & 4 & 1.38 & 2746 ($z=0.811$) & DESI ELG \cite{Gao2022}\\
    2 & $1.0 < z < 1.6$ & 6 & 3.24 & 2181 ($z=1.272$) & HSC NB816 \cite{okumura2022a}\\
    3 & $1.6 < z < 2.4$ & 3 & 5.29 & 1631 ($z=2.023$) & HSC NB912 \cite{okumura2022a}\\
    \hline
    \end{tabular}
    \caption{The detailed information of the three redshift bins. The first column shows the redshift range; the second column shows the mean galaxy number density in this redshift range according to PFS survey; the third column shows the comoving volume of the redshift bin; the fourth column shows the snapshot in the CosmicGrowth simulation used to conduct the mock kSZ map, galaxy and galaxy group catalogs in the redshift bin; the fifth column show the HOD model used to generate the ELGs. We assume the density and velocity baryon is following those of dark matter.}
    \label{Table:PFS_parameter}
    \end{table*}

\subsection{CosmicGrowth}
    The CosmicGrowth simulations \cite{Jing2019} are a series of high-resolution N-body cosmological simulations designed primarily to study the acceleration of cosmic expansion and the clustering of dark matter and galaxies. 
    The simulations are evolved with an adaptive parallel P3M -body code, while the force between the particles is softened at small scale. 
    In this work, we choose to use the “WMAP\_3072\_1200” simulation, whose box length is $1200\ {\rm Mpc}/h$.
    This simulation adopts a $\Lambda \rm CDM$ cosmological model with the cosmological parameters from WMAP satellite, $\Omega_b=0.0445$, $\Omega_c = 0.2235$, $\Omega_\Lambda = 0.732$, $h=0.71$, $n_s=0.968$ and $\sigma_8=0.83$. 
    It employs $3072^3$ dark matter particles. 
    The dark matter groups are identified by the Friends-of-Friends algorithm and the linking length is set to be 0.2 times the mean particle separation. A FoF group catalog is constructed for each snapshot, providing group mass, positions and peculiar velocities.
    Specifically, we use the snapshots 2746 ($z=0.811$), 2181  ($z=1.272$) and 1631  ($z=2.023$) to generate mock kSZ map, galaxy and galaxy group catalogs in the three redshift bins. 
        
\subsection{Mock kSZ/CMB maps}\label{subsec:mock_kSZ}
    \begin{figure}
        \centering
        \includegraphics[width=0.8\textwidth]{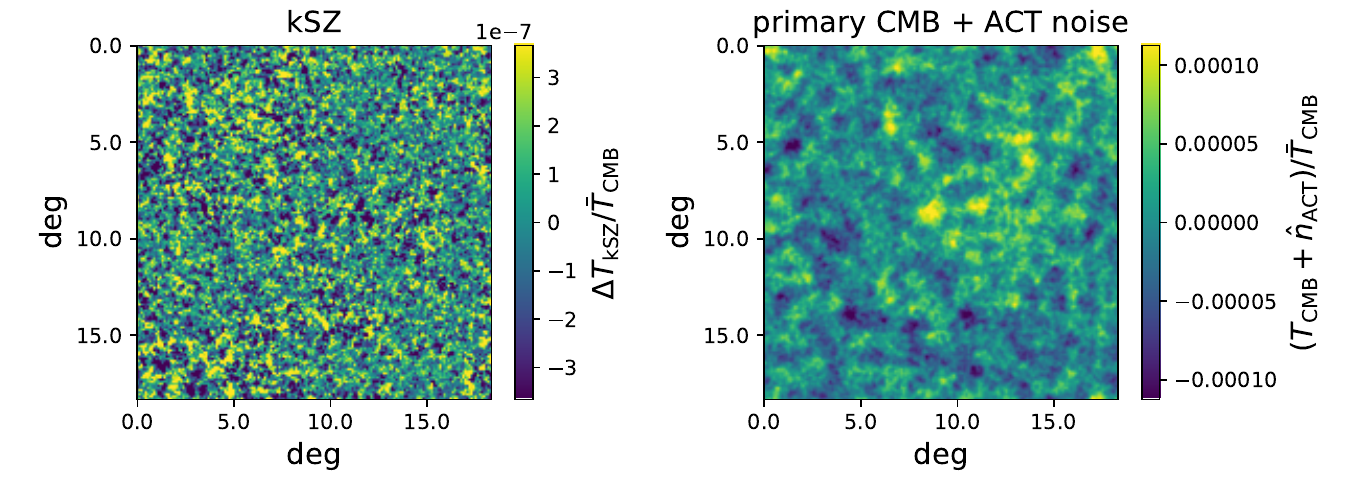}\\
        \includegraphics[width=0.8\textwidth]{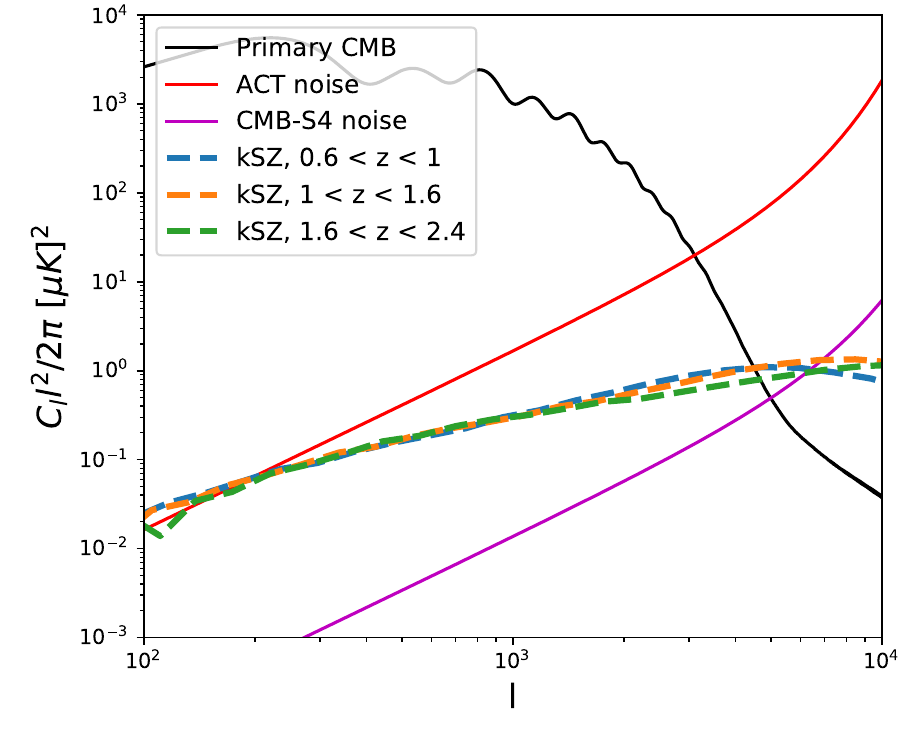}
        \caption{The top two panels show the mock sky map of the kSZ in redshift range $1.6 < z < 2.4$ (top left) and primary CMB with ACT noise (top right). The bottom panel shows the angular power spectrum of primary CMB (black), the instrument noise of ACT (red) and CMB-S4 (pink) and the angular power spectrum of the kSZ between $0.6 < z < 1$, $1 < z < 1.6$, $1.6 < z < 2.4$ (dashed lines).}
        \label{fig:kSZ_mock}
    \end{figure}
    %kSZ
    The amplitude of the kSZ effect between redshift $z_1$ and $z_2$ is the integrated momentum of electrons along the LOS direction
    \begin{eqnarray}
        \frac{\Delta T_{\rm kSZ}}{T_{\rm CMB}} = \int^{z_2}_{z_1} \chi_e\bar{n}_e(z)\sigma_T \frac{(1+\delta_e)\mathbf{v}\cdot\hat{n}}{c}e^{-\tau(z)}ad\chi,
    \end{eqnarray}
    where $\chi_e$ is the ionized baryon fraction, $\bar n_e(z)$ is the mean electron density, $\sigma_T$ is the Thomson scattering cross-section, $\delta_e$ and $\mathbf{v}$ is the over-density and peculiar velocity of electron, $\hat n$ is the LOS direction, $c$ is the speed of light, $\tau(z)=\int^z_0 \chi_e\bar{n}_e(z)\sigma_T ad\chi$ is the Thompson optical depth and $\chi$ is the comoving distance.
    We adopt $\chi_e = 1$, postulating complete ionization within our redshift interval of interest. 
    We further posit that the distribution and velocity of electrons follow those of dark matter, $\delta_{e} = \delta_{\rm dm}$ and $\mathbf{v}_{e} = \mathbf{v}_{\rm dm}$. While the baryonic processes would cause the divergence between free electrons and dark matter, we discuss in appendix~\ref{difference_dm_baryon} that this divergence is mostly diminished at large scale and may not cause a notable impact on our conclusion. With the position and velocity of dark matter particles from the simulation, we assign the momentum $\mathbf{p} \equiv (1+\delta_{\rm dm})\textbf{v}_{dm}$ to the $\rm GRID^3$ meshes. In this work, we set $\rm GRID = 2048$.
    By choosing an arbitrary LOS direction and integrating along this vector under the flat-sky approximation, we synthesize a simulated kSZ map. The redshift evolution in one redshift bin has not been considered here.
    
    %CMB+noise
    To construct maps for the primary CMB and the instrument noise, we utilize their angular power spectrum. 
    First, we establish the same two-dimensional gridding with that of the kSZ map in Fourier space. 
    Each point in this Fourier grid is endowed with a complex amplitude of the square root of the angular power spectrum $\sqrt{C_{l,X}}$, $X$ symbolizing either the CMB or noise, and is complemented by a randomly assigned phase. 
    Transforming the map from Fourier space to real space, a map of temperature fluctuations is achieved.
    Employing the CAMB software \cite{CAMB}, we simulate the primary CMB angular power spectrum, while the noise power spectrum is generated based on the instrumental white noise level $\Delta_T$
    \begin{eqnarray}
        C^{\rm Noise}_l = (\frac{\Delta_T}{T_{\rm CMB}})^2.
    \end{eqnarray}
    In addition we add a beam function 
    \begin{eqnarray}
        W_{\rm beam}(l) = \exp(-l(l+1)/l_{\rm beam}^2/2)
    \end{eqnarray}
    where $l_{\rm beam} = \sqrt{8\ln2}/\theta_{\rm FWHM}$ and $\theta_{\ \rm FWHM}$ is the beam size, to the kSZ and primary CMB components.
    The CMB instrument noise parameters are set to $\Delta_T = 11 \mu K - {\rm arcmin}$, $\theta_{\rm FWHM}=1.26 {\rm arcmin}$ for ACT case and $\Delta_T = 1 \mu K - {\rm arcmin}$, $\theta_{\rm FWHM}=1\ {\rm arcmin}$ for CMB-S4 case.
    In top panels of figure~\ref{fig:kSZ_mock}, we illustrate the mock kSZ and the primary CMB plus ACT instrumental noise map, to show the relative scale and amplitude of kSZ effect and its measurement noise, in the redshift interval $1.6 < z < 2.4$ as an example. The bottom panel shows the angular power spectra of kSZ effect in three redshift bins, $0.6 < z < 1.0$, $1.0 < z < 1.6$, $1.6 < z < 2.4$, primary CMB and the instrument noise for both ACT and the upcoming CMB-S4 experiment ($C^{\rm Noise}_l/W^2_{\rm beam}(l)$). 
    Notably, the amplitude of the kSZ effect appears lower than the combined intensity of the CMB and noise, by more than two orders of magnitude. This underscores the challenge of distinguishing the kSZ signal against the background of primary CMB fluctuations and instrumental noise, emphasizing the need for robust techniques to extract and interpret the kSZ signal accurately.
    
\subsection{Mock galaxy, galaxy group catalogs}\label{subsec:mock_catalog}
    We generate the galaxy catalog from the FOF group catalog from the  simulation and the ELG HOD. In this work we focus on the redshift range $0.6< z< 2.4$. The redshift range is divided into three redshift bins, $0.6 <z <1$, $1< z<1.6$, $1.6< z<2.4$, of which we use different ELG HODs (the details are shown in appendix~\ref{app:HOD}). To generate the galaxy catalog mock, we use the following steps:
    \begin{itemize}
        \item 
        Using the HOD to determine whether there is a central galaxy and how many satellite galaxies in each dark matter halo and generating a galaxy catalog.
        \item 
        Randomly choosing galaxies to make number density of the catalog consistent with the value from PFS science strategy paper (list in table~\ref{Table:PFS_parameter}).
        \item 
        Giving the galaxies the same position and peculiar velocity as their host halo. Here, we ignore the relative distance and velocity between a galaxies and its host halo.
        \item 
        Assigning the RSD to LOS direction position of galaxies.
    \end{itemize}
    We use this spectroscopic redshift galaxy mock catalog to reconstruct velocity with the method presented in section.~\ref{sec:vel_rec_method}.\\
    
    For the galaxy group catalog, we use the halos from the simulation whose mass is above a specific mass limitation. The photometric redshift uncertainty is assigned by $\sigma_z = 0.01(1+z)$. We use this galaxy group catalog to determine where to stacking the filtered CMB temperature. This galaxy group sample is assumed to be identified from photometric galaxies using a group finder such as ref.~\cite{Yang2007}. The galaxy groups can point out the positions where is abundant of free electron and improve the measurement significance.

\section{Forecast kSZ measurements with PFS survey}\label{sec:forecast}
    \begin{figure*}
	\includegraphics[width=1\textwidth]{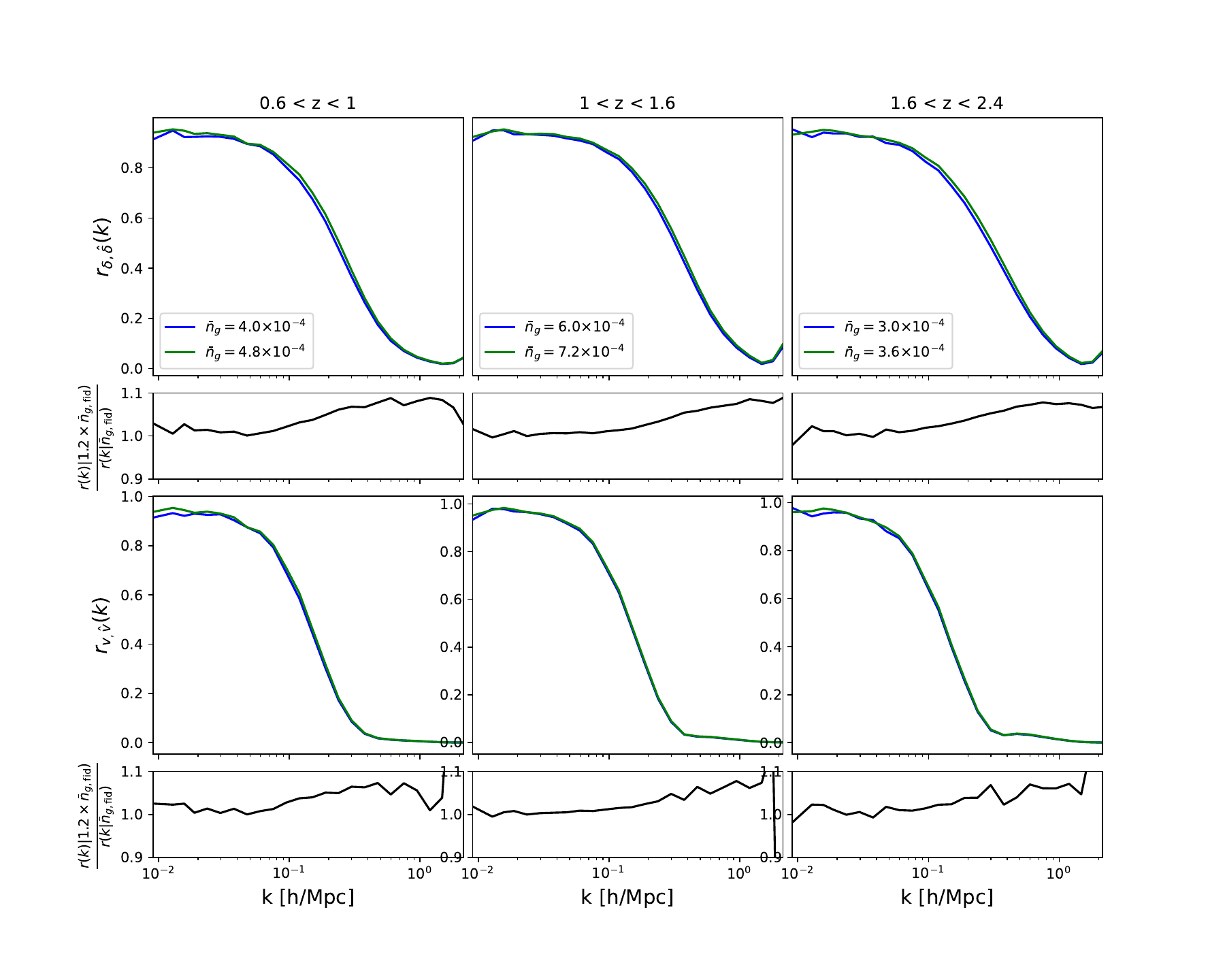}
    \caption{The correlation functions of the reconstructed density and velocity fields with the true fields from N-body simulation in three redshift bins. The first row shows the correlation function of the reconstructed density and the matter density from simulation. The blue lines represent the density reconstructed from the spectroscopic galaxy catalog with a fiducial PFS galaxy number density, while the green lines represent the density reconstruction with 20\% more galaxies. 
    The second row shows the ratio of the green and blue line, elucidating how an augmentation in galaxy number density can potentially enhance the performance of reconstruction. 
    The third and forth rows shows the same analysis but focus on the reconstructed velocity field. 
    For comparison, the correlation parameter between the reconstructed velocity and the true values of individual galaxy groups is from 0.31 to 0.38, depending on the galaxy group mass and redshift.\label{fig:rk}}
    \end{figure*}
    
    In this section, we explore the efficacy of the stacking and tomography methods by assessing the kSZ effect measurement significance, utilizing the PFS survey. Both methods incorporate a crucial step, the reconstruction of velocity, that significantly impacts measurement accuracy. To quantify the performance of the velocity reconstruction, we employ the correlation function
    \begin{eqnarray}
        r(\mathbf k) = \frac{\langle {v_{\rm los}}^{\rm True}(\mathbf k) {v_{\rm los}}^{*,\rm Rec}(\mathbf k)\rangle}{\langle {v_{\rm los}}^{\rm True}(\mathbf k) {v_{\rm los}}^{*,\rm True}(\mathbf k)\rangle\langle {v_{\rm los}}^{\rm Rec}(\mathbf k) {v_{\rm los}}^{*, \rm Rec}(\mathbf k)\rangle}.
    \end{eqnarray}
    And the correlation parameter of the galaxy groups' LOS velocity  
    \begin{eqnarray}
        r = \frac{\langle {v_{c,\rm los}}^{\rm True} {v_{c,\rm los}}^{*,\rm Rec}\rangle}{\langle {v_{c,\rm los}}^{\rm True} {v_{c,\rm los}}^{*,\rm True}\rangle\langle {v_{c,\rm los}}^{\rm Rec} {v_{c,\rm los}}^{*, \rm Rec}\rangle}.
    \end{eqnarray}
    The bottom panels of figure~\ref{fig:rk} reveal the correlation function between the peculiar velocity field of dark matter particles and the reconstructed velocities in three distinct redshift bins. Notably, the correlation parameter remains high ($> 0.9$) at larger scales $k<0.03\ h/{\rm Mpc}$ but it rapidly diminishes, approaching zero at smaller scales  $k > 0.3\  h/{\rm Mpc}$. Furthermore, an increase in the fiducial galaxy number density by 20\% yields a scale-dependent improvement in the correlation function. This improvement is minimal at $k<0.07\ h/{\rm Mpc}$ and increases by about 7\% for higher $k$ value.
    
\subsection{Stacking method}
    \begin{figure}
    \centering
	   \includegraphics[width=0.7\textwidth]{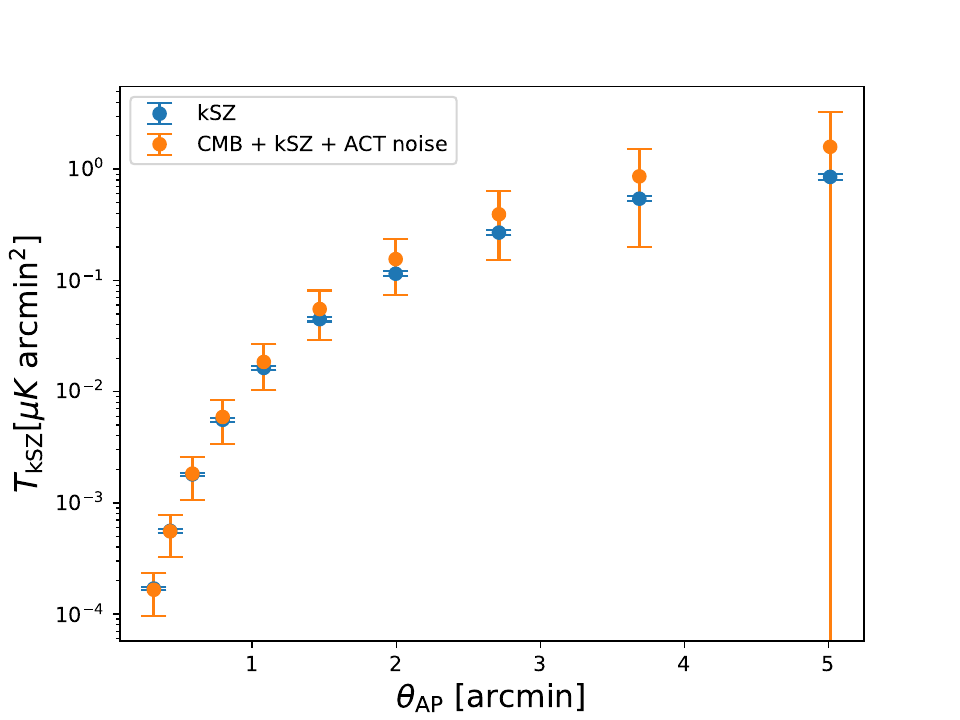}
	   \caption{The kSZ profile from the stacking measurement (eq.~\ref{eq:stack_observable}). The galaxy group mass range is $12.29 < {\rm lgM} < 12.39$ and the redshift range is $1.6 < z < 2.4$. The blue points show the case with only kSZ component and the orange points show the case with primary CMB, kSZ and ACT instrument noise components. \label{fig:stacking_signal}}
    \end{figure}
    \begin{figure*}
	   \includegraphics[width=1\textwidth]{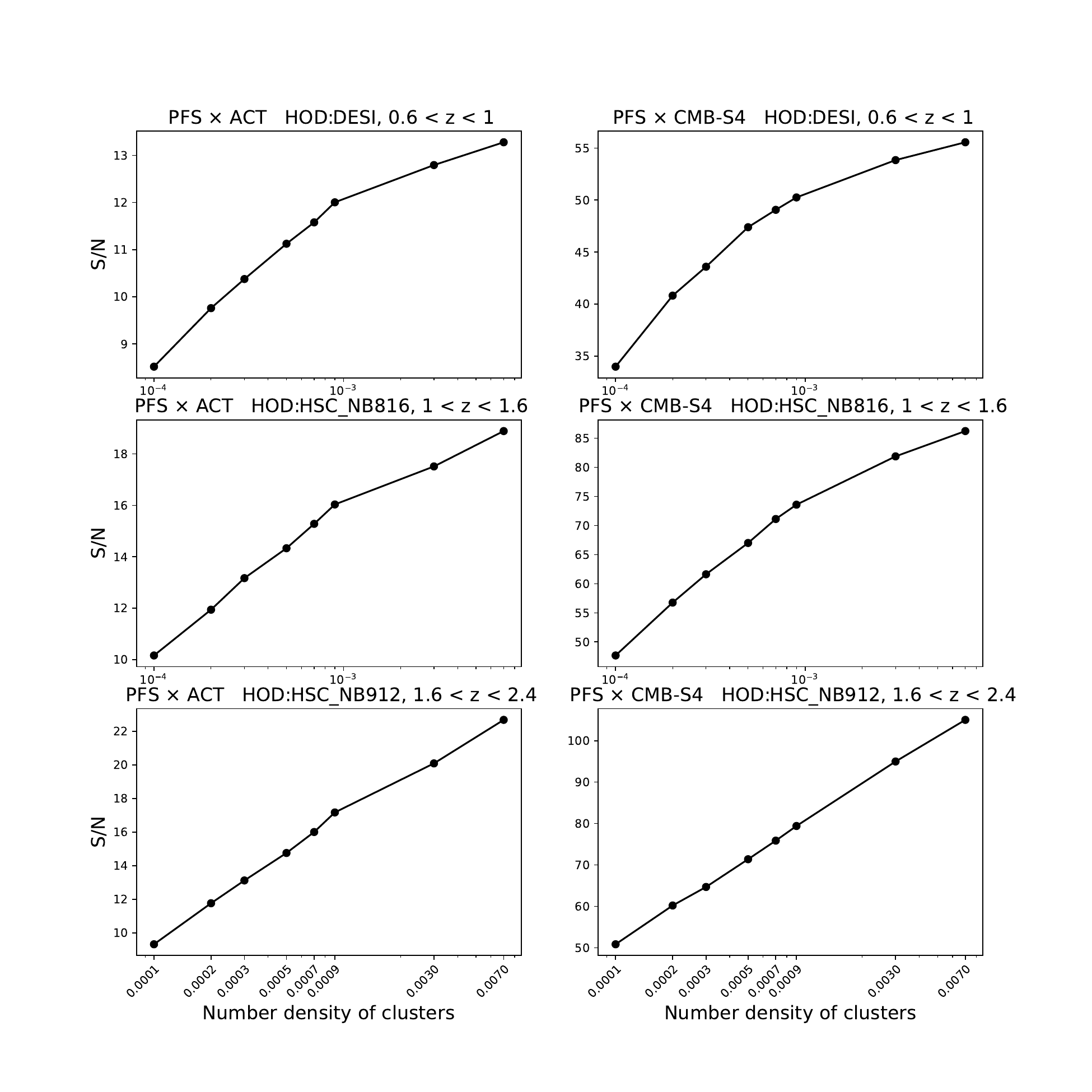}
	   \caption{The measurement significance of the kSZ stacking method. The three panels from top to bottom represent the redshift bins $0.6 < z < 1.0$, $1.0 < z < 1.6$ and $1.6 < z < 2.4$ for the combination of PSF $\times$ ACT (left) and PSF $\times$ CMB-S4 (right). \label{fig:stack_SN}}
    \end{figure*}
    The kSZ stacking method involves the stacking of the filtered CMB temperature signals at the position of galaxy groups weighted by their LOS velocity. The LOS velocity is reconstructed using the method outlined in section~\ref{sec:vel_rec_method} and a spectroscopic redshift galaxy catalog, as described in section~\ref{subsec:mock_catalog}. This velocity-weighted stacking helps suppress the tSZ effect, dust contamination, and other foregrounds that do not correlate with the LOS velocities of the galaxy groups.

    In the first place, to suppress the instrument noise, we add a matched filter
    \begin{eqnarray}
        W_{\rm matched}(l) = \sqrt{\frac{C_{\rm CMB}(l)+C_{\rm kSZ}(l)}{C_{\rm CMB}(l)+C_{\rm kSZ}(l)+C_{\rm instr}(l)}}.
    \end{eqnarray}
    Additionally, the amplitude of the power spectrum of the primary CMB is larger than that of the kSZ effect by more than three orders of magnitude. In larger scales, the CMB component is more dominant, and at the scale of galaxy groups, it exhibits almost none variation. To effectively isolate the kSZ signal from the dominating CMB fluctuations, an aperture photometry (AP) filter is applied at the location of each galaxy group
    \begin{eqnarray}
        T_{{\rm AP},i}(\theta_{\rm AP})= \int T(\Vec{\theta}-\Vec{\theta_i})W_{\rm AP}(\theta)d\Vec{\theta}, 
    \end{eqnarray}
    where $\Vec{\theta_i}$ is the angular position of the galaxy group $i$.
    The AP filter
    \begin{eqnarray}
	W_{\rm AP}(\theta)=\frac{1}{\pi\theta^2_{\rm AP}}
	\left\{
	\begin{array}{cc}
		1,&\quad \theta\leq\theta_{\rm AP}  \\
		-1,&\quad \theta_{\rm AP} < \theta \leq \sqrt{2}\theta_{\rm AP}\\
		0,&\quad \theta > \sqrt{2}\theta_{\rm AP} 
	\end{array}
	\right.
    \end{eqnarray}
    designed to be compensated, integrates over an area to zero. This feature enables the filter to cancel out fluctuations that have a wavelength longer than its size.
    Note if the size of the AP filter is smaller or comparable to the size of the galaxy groups, the AP filter would subtract the kSZ signal in the annulus region, thereby suppressing the total kSZ signal \citep{soergel2018a, gong2023a}.
    By varying the size of the AP filter from 0.3 to 5 arcmin, it is possible to delineate the kSZ profile around galaxy groups. The stacking estimator is 
    \begin{eqnarray}
    \label{eq:stack_observable}
        T_{\rm kSZ}(\theta_{\rm AP}) = \frac{1}{r_v}\frac{v^{\rm rms}_{\rm rec}}{c} \frac{\sum_i T_{{\rm AP},i}(\theta_{\rm AP}) (v_{{\rm rec}, i}/c)}{\sum (v_{{\rm rec}, i}/c)^2},
    \end{eqnarray}
    where $T_{{\rm AP},i}(\theta_{\rm AP})$ is the CMB temperature at the position of the galaxy group $i$ after applying the AP filter whose size is $\theta_{\rm AP}$, $v_{{\rm rec}, i}$ is the reconstructed LOS velocity of the galaxy group $i$, $v^{\rm rms}_{\rm rec}$ is the root mean square of the reconstructed LOS velocity. And $r_v$ is the correlation coefficient of the reconstructed and true los galaxy group velocity.\\

    We use the Jackknife method \cite{Hartlap2007} to estimate the precision of measurements. The galaxy group catalog is divided into $N_{\rm JK}$ subsamples. Then, we measure the observable (eq.~\ref{eq:stack_observable}) by removing one subsample from the dataset at a time. The Jackknife covariance is 
    \begin{eqnarray}
        C^{\rm JK}_{ij} = \frac{N_{\rm JK}-N_{\rm bins}-2}{N_{\rm JK}-1} \sum_k^{N_{\rm JK}} (T^k_{\rm kSZ}(\theta_{\rm AP, i})-\bar T_{\rm kSZ}(\theta_{\rm AP, i})) 
        \nonumber \\ 
        \times (T^k_{\rm kSZ}(\theta_{\rm AP, j})-\bar T_{\rm kSZ}(\theta_{\rm AP, j})),
    \end{eqnarray}
    where $k$ represents each measurement, $i$ and $j$ represent the $i$-th ans $j$-th bin for the AP filter size, $\bar T_{\rm kSZ}$ is the mean value of $N_{\rm JK}$ measurements and $N_{\rm bin}$ is the number of the AP filter sizes.

    The S/N is estimated by 
    \begin{eqnarray}
        \frac S N = \sqrt{\frac{V_{\rm survey}}{V_{\rm simu}}\cdot\chi^2_{\rm null}},
    \end{eqnarray}
    where $V_{\rm survey}$ is the comoving survey volume for the redshift bin and $V_{\rm simu} = 1200^3\ [{\rm Mpc}/h]^3$ is the simulation volume, and 
    \begin{eqnarray}
        \chi^2_{\rm null} = \sum_{ij}\bar T_{\rm kSZ}(\theta_{\rm AP, i}) \left(C^{{\rm JK}}_{ij}\right)^{ -1}\bar T_{\rm kSZ}(\theta_{\rm AP, j}).
    \end{eqnarray}
    
    %results

    Figure~\ref{fig:stacking_signal} show the kSZ profile measurement of a galaxy group sample whose mass range is $12.29 < {\rm lgM} < 12.39$ in the third redshift bin with the ACT survey as an example. The kSZ-only case is consistent with the one adding primary CMB and ACT instrument noise within the errorbars, showing the stacking estimate (eq. \ref{eq:stack_observable}) is unbiased. Figure~\ref{fig:stack_SN} presents our forecast of the kSZ stacking measurement significance across three redshift bins for both the ACT and CMB-S4 surveys. Notably, the CMB-S4 survey offers a significant improvement, approximately a fivefold enhancement compared to the ACT survey. This improvement arises primarily from the superiority sensitivity of CMB-S4. When compared to the ACT case, the stacking profiles from CMB-S4 exhibit a higher S/N for $\theta_{AP}<2$ arcmin. However, the characteristic scale of galaxy groups ($\sim 1$ arcmin, corresponding to $l\sim 2\times 10^{4}$) is dominated by instrumental noise for both ACT and CMB-S4.

    \begin{table*}
    \centering
    \begin{tabular}{c|cccccccc}
    \hline
     & \multicolumn{8}{c}{$\bar n_g \times 10^{-4} [{\rm Mpc}/h]^{-3}$} \\
     \hline
    Redshift bin & 1 & 2 & 3 & 5 & 7 & 9 & 30 & 70 \\
    \hline
     $0.6 < z < 1.0$ & 13.30 & 13.08 & 12.94 & 12.75 & 12.62 & 12.52 & 12.02 & 11.65\\
     $1.0 < z < 1.6$ & 13.13 & 12.92 & 12.80 & 12.62 & 12.50 & 12.41 & 11.95 & 11.60\\
     $1.6 < z < 2.4$ & 12.84 & 12.67 & 12.54 & 12.39 & 12.29 & 12.21 & 11.80 & 11.49\\
    \hline
    \end{tabular}
    \caption{The minimum masses of galaxy group samples in the three redshift bins. Each column corresponds to a galaxy group number density, as labeled in the second line.}
    \label{Table:minimum_mass}
    \end{table*}

    Our findings further reveal a notable dependence of the S/N on galaxy group number density. Specifically, increasing the galaxy group number density within the range of $10^{-4}$ to $7\times 10^{-3}\  [h/{\rm Mpc}]^3$ leads to an approximately twofold increase in the S/N. The corresponding minimum masses of these galaxy group samples are shown in table \ref{Table:minimum_mass}. Interestingly, augmenting galaxy density by 20\% yielded negligible changes in measurement significance and the correlation coefficient. This aligns with the findings of ref.~\cite{guachalla2023a}, suggesting that the reconstruction performance saturates when the galaxy number density exceeds $5\times 10^{-5}[h/{\rm Mpc}]^3$.
    
    Furthermore, the correlation parameter between true and reconstructed galaxy group LOS velocity is between 0.31 to 0.38, and we observe a decreasing trend in the correlation parameter with increasing galaxy group mass (in this case, inferred from photometric galaxy group redshifts). This relationship is reversed if the galaxy group redshift is spectroscopic instead of photometric. Notably, the LOS velocity amplitude of galaxy groups exhibits no apparent dependence on galaxy group mass. We therefore conclude that the decreasing correlation for massive galaxy groups arises from the large velocity disparity around them.
    
    In addition, if the galaxy group number density could reach $1\times 10^{-3}\  [h/{\rm Mpc}]^3$, the S/N of the kSZ stacking method for the three redshift bins would be expected to reach 12, 17, and 18 for the ACT case and 52, 75, and 85 for the CMB-S4 case, respectively. Two primary factors influence the stacking measurements in different redshift bins: survey volume and galaxy group angular radius. For a fixed galaxy group number density, a larger survey volume translates to a higher S/N. However, at higher redshifts, the galaxy groups with same mass will have a smaller angular radius, making it more susceptible to instrumental noise.

    Finally, in this prediction analysis, we use an external photometric galaxy group catalog as the position proxies for stacking kSZ effect. However, the certain galaxies, such as LRGs, can serves as proxies of galaxy groups. Thus, the full stacking measurement could just use the PFS LRG catalog. We leave this to future work.

\subsection{Tomography method}
    \begin{figure}
        \centering
	   \includegraphics[width=0.5\textwidth]{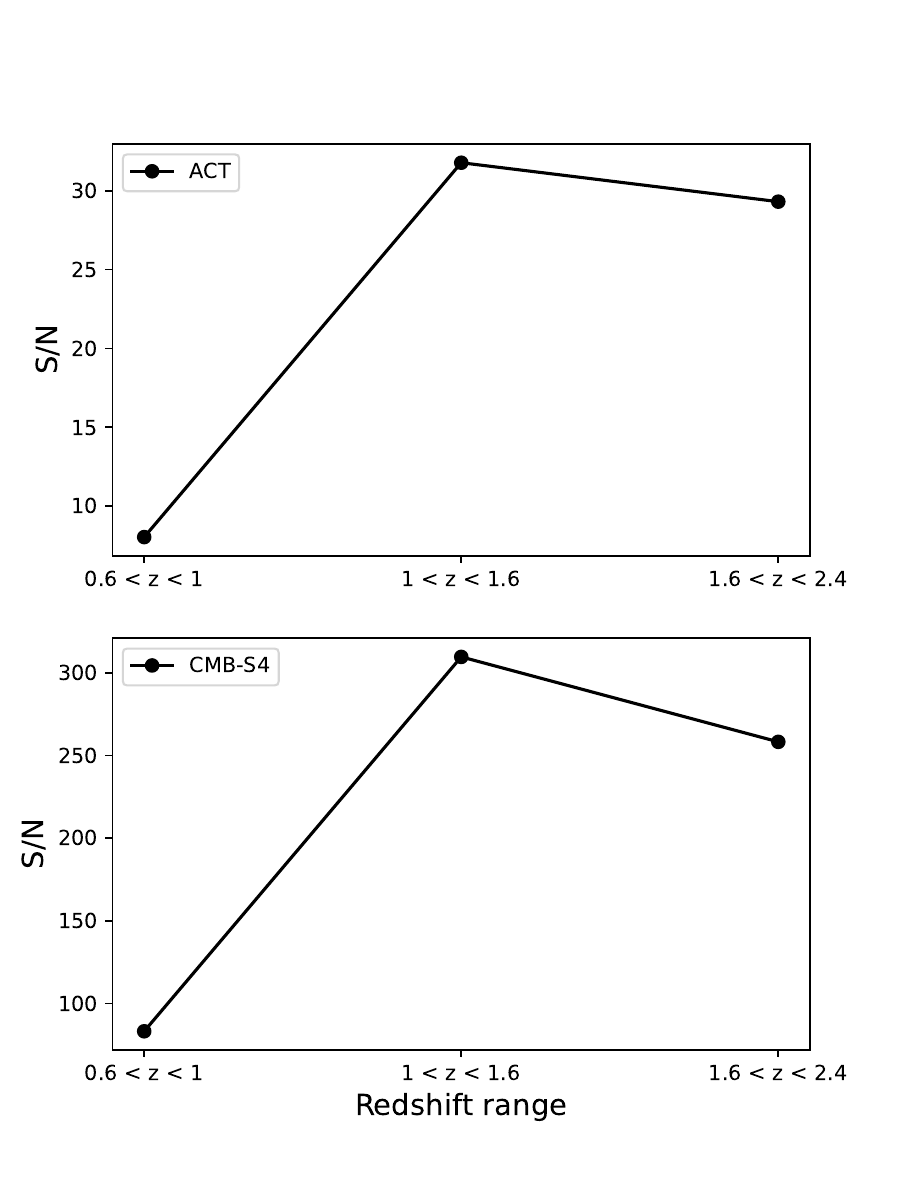}
	   \caption{The measurement significance of the kSZ tomography method for the combination of PSF $\times$ ACT (top) and PSF $\times$ CMB-S4 (bottom). The x-axis shows the redshift range of the bins. The y-axis shows the S/N of the measurements.  \label{fig:tomography_SN}}
    \end{figure}

    We follow the steps in ref.~\cite{shao2011a} to measure the kSZ effect by the tomography method. A kSZ template is constructed by integrating
    \begin{eqnarray}
    \label{eq:kSZ_tomography_template}
        \hat \Theta (\hat{n})= \int^{z_2}_{z_1}W(z)\frac{\hat{\mathbf{p}}(\mathbf{\chi \hat{n}, \chi})\cdot \hat{n}}{c} ad\chi,
    \end{eqnarray}
    where $\hat{n}$ is the LOS direction, $\chi$ is the comoving distance, $a$ is the scale factor, $c$ is the speed of light, $z$ is the redshift , $W(z)=\chi_e\bar n_e(z)\sigma_Te^{-\tau(z)}$ is the redshift weight. In the reconstructed momentum $\hat{\mathbf{p}} = \hat{\mathbf{v}} \cdot \hat \delta$, $\hat \delta$ is the galaxy density field with the weiner filter $W_2(k)$ and $\hat{\mathbf{v}}$ is the reconstructed velocity described in section~\ref{sec:vel_rec_method}. 
    Then, correlate it with the CMB survey maps. The statistical error of the correlation angular power spectrum $C_l$ is 
    \begin{eqnarray}
        \left(\frac{\Delta C_l}{C_l}\right)^2 = \frac{1}{2l\Delta l f_{\rm sky}}\frac{(C^{\rm CMB}_l+C^{\rm noise}_l)C^{\hat\Theta\hat\Theta}_l}{(C^{\Theta\hat\Theta}_l)^2},
    \end{eqnarray}
    where $\Theta$ is the kSZ signal $\Delta T_{\rm kSZ}/T_{\rm CMB}$ in this redshift bin, $f_{\rm sky}$ is the sky fraction, $C^{\rm CMB}_l$ is the angular power spectrum of the primary CMB power spectrum and $C^{\rm noise}_l$ is the angular power spectrum of noise. 
    The total measurement significance of the tomography can be estimated by 
    \begin{eqnarray}
        \frac{S}{N}=\frac 1 {\sqrt{\sum \left({\Delta C_l}/{C_l}\right)^2}}.
    \end{eqnarray}

    Figure \ref{fig:tomography_SN} depicts the significance of the kSZ tomography measurement, quantified by the S/N, across three redshift bins ($0.6 < z < 1$, $1 < z < 1.6$, and $1.6 < z < 2.4$) for both the ACT and CMB-S4 surveys. The S/N values for the ACT survey are 8, 32, and 28, while the CMB-S4 survey achieves significantly higher values of 80, 310, and 250, respectively. This roughly ten-fold improvement highlights the superior of CMB-S4 for kSZ tomography measurements.

    The relatively low S/N observed for the ACT survey in the first redshift bin ($0.6 < z < 1$) suggests limitations in current CMB surveys for measuring the kSZ effect via tomography at redshifts below z = 1. In contrast, the S/N ratio exhibits a significant rise for both surveys in the second and the third redshift bins.
    
    This improvement is attributed to several factors. 
    Firstly, at higher redshifts, the comoving number density of free electrons increases, leading to a stronger kSZ signal.
    Secondly, larger survey volumes encompass a greater number of galaxy groups, contributing to a higher S/N.
    Thirdly, for the second redshift bin, while a higher galaxy density may not significantly improve velocity reconstruction, it does contribute to a more accurate reconstruction of the underlying density field, which is crucial for tomography measurements.
    Fourthly, the non-linear components are reduced with the increasing of the redshift and this would also increase the reconstruction performance.
    
    In addition, we find a 20\% increase in galaxy number density is projected to improve the S/N ratio by approximately 10\% across all redshift bins, underscoring the substantial impact of galaxy density on the correlation between the kSZ signal and the reconstructed template.
    Incorporating complementary tracers of large-scale structure, beyond galaxy surveys alone, could potentially mitigate the impact of Poisson noise and further enhance the significance of measurements at higher redshifts.\\

    Above all, both stacking and tomography methods in the PFS survey demonstrate powerful performance in high redshift bins. This indicates a significant potential for the PFS survey in measuring the kSZ effect at high redshifts, marking it as a key merit of this observational program. While the kSZ stacking and tomography estimators are theoretically equivalent to the bispectrum $\langle \delta\delta v \rangle$ (appendix~\ref{appkSZMethodEquivalence}), they measure the 1h-term and 2h-term components respectively, like correlation function and power spectrum.    
    Moreover, the kSZ tomography measurement only relying on PFS spectroscopic galaxy catalog possesses higher S/N than that of stacking method in the two high redshift bins. This is attributed to the tomography signal capturing contributions from free electrons across various cosmic structures, including halos, filaments, and voids, whereas the stacking approach predominantly measures kSZ effects from massive clusters. Additionally, the scale region where the 1h-term dominates ($l \sim 10^4$) is severely suffered by instrumental noise, even within the CMB-S4 survey. 
    In addition, the kSZ tomography is more sensitive to the galaxy number density compared to stacking method. The increasing of galaxy number density would not only improve the velocity reconstruction performance, but also the reconstructed density field itself. We find the improvement of the correlation of the reconstructed and true field shows up only when $k>0.03 h/ {\rm Mpc}$ and slightly increases with the increasing of $k$ (figure.~\ref{fig:rk}). Therefore, the tomography method benefits from both improvement of small scales and reconstructed density fields. 
    Lastly, the instrument noise of CMB-S4 is about ten times smaller than that of ACT. Therefore, for both stacking and tomography methods, the S/N for CMB-S4 is  significantly higher than that of ACT.

\section{Conclusion}\label{sec:conclusion}
    In this study, we explore the measurement significance of applying kSZ stacking and tomography to the PFS spectroscopic redshift galaxy survey and CMB surveys like ACT and CMB-S4. Both methods involve velocity reconstruction utilizing the density field from a biased tracer, specifically the PFS spectroscopic galaxies. 
    In the velocity reconstruction, the noise of density field caused by the RSD and shot noise should be suppressed at small scale. Compared to the Gaussian filter, our findings indicate that the optimal Wiener filter slightly enhances the correlation between the true and reconstructed velocity for $k>0.1 h/{\rm Mpc}$. 
    
    To forecast the measurement significance, we use an N-body simulation to generate mock kSZ map, galaxy and group catalogs. For the kSZ mock, we assume the baryon distribution trace that of dark matter. 
    This assumption does not account for the redistribution of baryons driven by baryonic processes. These processes, such as supernovae and AGN feedback, can expel baryonic material from galaxy groups into the surrounding environment. This effect is likely a subdominant contributor to forecasting kSZ measurement significance and will be addressed in future work. 
    Here, we just discuss how the baryon feedback would influence our results. The stacking method is more sensitive to the feedback especially when $\theta_{\rm AP}$ is small, because the signal of kSZ effect in groups is proportional to the amount of baryon. Therefore, the S/N is decreasing with a scaling of $f_{\rm gas}$. On the other hand, the stacking measurement and its S/N can provide a constraint on $f_{\rm gas}$ and then the baryon feedback model.

    To generate the realistic PFS galaxy catalog, we employ realistic ELG HODs from a dark halo catalog. Randomly selecting galaxies ensures the galaxy number density aligns with the predicted value from PFS. Our results underscore the advantages of high redshift kSZ measurement due to its substantial signal amplitude and better velocity reconstruction performance in both methods.
    In the stacking method, while using the PFS galaxy sample to reconstruct velocity, we use another galaxy group sample to pin down where to stack kSZ temperature.  Interestingly, we observe that the S/N is strongly dependent on the number density of the galaxy group sample rather than the density of PFS galaxies. The tomography method, which only needs the PFS galaxy sample, can detect kSZ effect with the S/N = 8, 32, 28 for ACT survey and 80, 310, 250 for CMB-S4 survey respectively in the three redshift bins $0.6 < z < 1.0$, $1.0 < z < 1.6$ and $1.6 < z < 2.4$.
    
    While the stacking method excels in measuring the detailed electron profile around massive galaxy groups, tomography captures all free electrons in groups, filaments, and voids, making it sensitive to the global baryon abundance. These two methods are complementary. Leveraging the PFS spectroscopic redshift survey, the kSZ measurement allows the constraint of baryon abundance extension to $z=2.4$. 
    Simultaneously, the kSZ measurement serves as a valuable tool to constrain baryonic process strength, offering insights into the distribution of baryons. This provides an observable means to calibrate sub-grid parameters in hydrodynamic simulations and address one of the most significant theoretical systematic uncertainties in weak lensing measurements.

    Furthermore, to enhance the measurement significance beyond redshift $z=1$, the incorporation of other high redshift data such as, the Lyman-$\alpha$ forest and 21cm intensity mapping, could significantly improve the reconstruction performance of peculiar velocities and redshift range. The observed increase in quasar number density facilitates more accessible and accurate density measurements through their absorption lines. But the sampling across the sky can be sparse. This characteristic poses challenges for velocity reconstruction, which we defer to future investigations.

\acknowledgments
This work made use of the Gravity Supercomputer at the Department of Astronomy, Shanghai Jiao Tong University. 
This work was supported by the National Key R\& D Program of China (2023YFA1607800,2023YFA1607801,2020YFC2201602), the China Manned Space Project (\#CMS-CSST-2021-A02), and the Fundamental Research Funds for the Central Universities.

\appendix
\section{The difference between dark matter and baryon}\label{difference_dm_baryon}
\begin{figure}
	\includegraphics[width=0.5\textwidth]{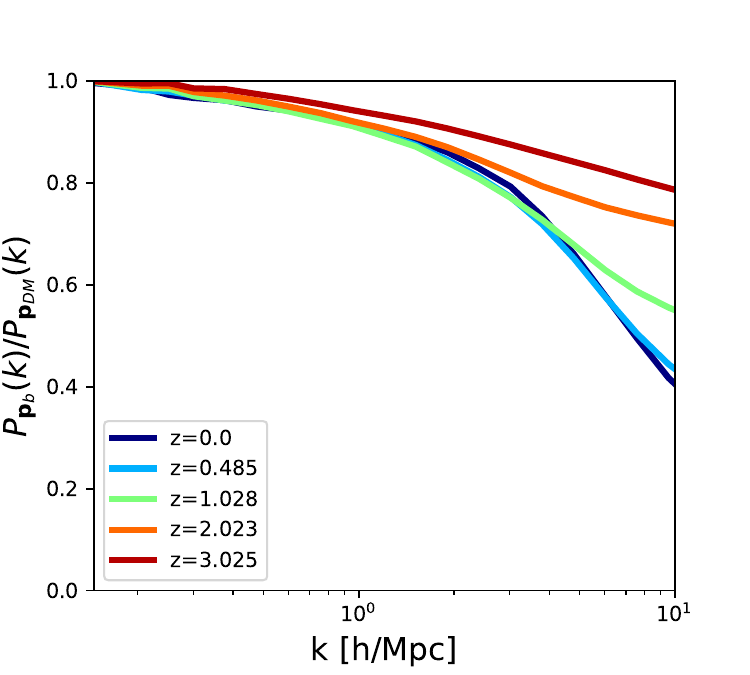}
    \includegraphics[width=0.5\textwidth]{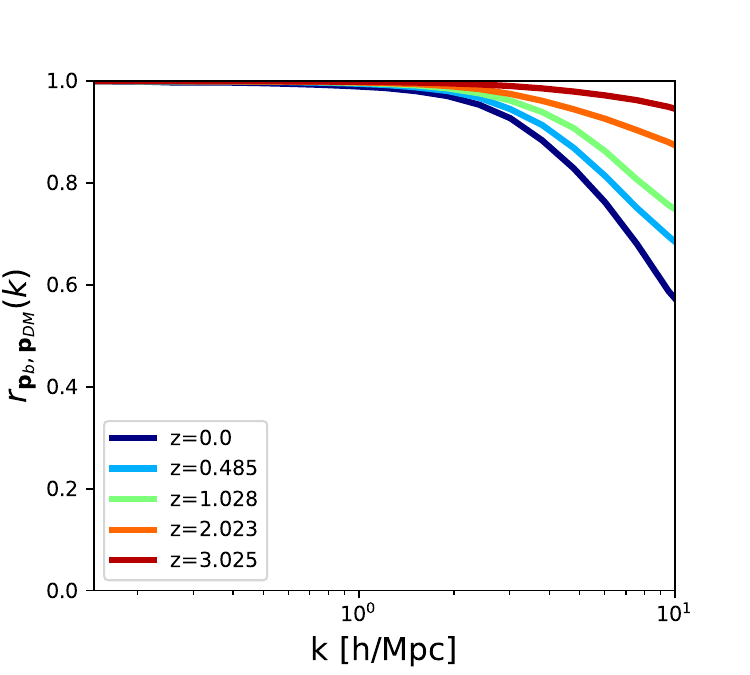}
	\caption{The ratio of baryon and dark matter momentum power spectrum (top) and the correlation function between them (bottom) from TNG100-1 simulation. \label{fig:diff}}
\end{figure}
In this study, we leverage dark matter data from N-body simulations to construct a kSZ mock. However, it is essential to acknowledge that the kSZ effect is intricately linked to the distribution and velocity of free electrons (baryons), which inherently differ from those of dark matter due to baryonic processes, such as AGN and Supernovae feedback. 

This section aims to elucidate the distinctions in momentum between dark matter and baryons, employing hydrodynamic simulations from TNG100-1, a subset of the IllustrisTNG project. IllustrisTNG is a suite of state of art hydrodynamic simulations designed to explore galaxy formation and evolution in the universe. The TNG100-1 simulation incorporates $2500^3$ dark matter particles and gas cells within a volume of $75^3 [{\rm Mpc}/h]^3$, adopting to a cosmological model from ref.~\cite{TNG_Planck}. Position and velocity data for dark matter particles and gas cells are easily retrieved from the simulation, with their momentum, $\mathbf{p} = (1+\delta)\mathbf{v}$, allocated to a grid of $512^3$. Then we calculate the power spectrum and correlation function of these two momentum fields and show them in figure~\ref{fig:diff}.

In the top panel, we present the ratio of the power spectrum of baryon momentum to that of dark matter. Notably, at large scales ($k<0.2 h/{\rm\ Mpc}$), the ratio approaches unity and diminishes with increasing wave number $k$. At $k=1 h/{\rm\ Mpc}$, the ratio remains above 0.9 for all redshifts. The bottom panel illustrates the correlation function between dark matter and baryon momentum, demonstrating near unity correlation at scales $k<1 h/{\rm\ Mpc}$ for all redshifts, with a decrease as $k$ increases.
Furthermore, both the ratio and correlation tend towards unity when $k<1 h/{\rm\ Mpc}$, indicating diminished differences between dark matter and baryons, especially at higher redshifts. Although, in the main body of our analysis, we assume that the distribution and velocity of baryons mimic those of dark matter. We acknowledge this assumption may cause potential bias in kSZ mock, particularly at lower redshifts and smaller scales. However, for the redshift and scale range of our interest, this bias is expected to have a negligible impact on the estimation of kSZ measurement significance.

\section{Connection between halos and galaxies}
\label{app:HOD}
    The CosmicGrowth is a dark-matter simulation which only contain halos and subhalos. In this work, we use the ELG sample from PFS to reconstruct the velocity field. The ELG is a type of galaxy characterized by their strong emission lines in their spectra. These emission lines are typically caused by gas within the galaxy being excited by young, hot stars or AGNs. ELGs often exhibit high levels of star formation activity. They are not necessary to be bright reside in the massive halos. We employ three kinds of ELG HOD to sample the galaxies into the halos. 
    The first is from \cite{Gao2022}. They use the galaxy catalog of the final public release of the VIPERS and a photometric catalog from Canada-France-Hawaii Telescope Legacy Survey. They choose the galaxy sample in the redshift range $0.5 \leq z \leq 0.8$ and divide the galaxies into four luminosity bins, and fit the HOD for each bin. The HOD model formula they used is 
    \begin{eqnarray}
    \label{eq:HOD_parameter}
        N_{\text{cen}}(M) &=& F^B_c \left(1 - F^A_c\right) \exp\left[ -\frac{\left(\log(M/M_c)\right)^2}{2\sigma^2_{\log M}} \right] 
        \\ \nonumber
        &&+ F^A_c \left[1 + \text{erf}\left(\frac{\log(M/M_c)}{\sigma_{\log M}}\right)\right] \times \left(1 + \frac{M}{M_c}\right)^{\beta_c}\\
        N_{\text{sat}}(M) &=& F_s \left[1 + \text{erf}\left(\frac{\log(M/M_{\text{min}})}{\delta_{\log M}}\right)\right] \times \left(\frac{M}{M_{\text{min}}}\right)^{\alpha_s}.
    \end{eqnarray}
    We used this HOD to sample the galaxy at the redshift range $0.6-1$ in the main text.
    And the second is from \cite{Teppei2021}, which use the the ELGs identified by two narrow-band, NB816 (z=1.19) and NB912 (z=1.47), in the HSC-SSP survey. They used the same HOD model formula with the first one, but $\beta_s = 0$ We use these two HODs for the redshift range $1.0\sim 1.6$ and $1.6\sim 2.4$ respectively. The value of the parameters in these HODs are listed in the table~\ref{Table:HOD_parameters}.

    \begin{table*}
    \label{Table:HOD_parameters}
    \centering
    \begin{tabular}{ccccccccccc}
    \hline
    Redshift bin & Sample &  $\log M_c $ & $\sigma_{\log M}$ & $F^A_c$ & $F^B_c$ & $\beta_c$ & $\log M_{\mathrm{min}} $ & $F_s$ & $\delta_{\log M}$ & $\alpha_s$ \\
    \hline
    \multirow{4}{*}{$0.6 \leq z \leq 1.0$} & L0 & 11.234 & 0.206 & 0.133 & 0.010 & -0.185 & 11.690 & 0.015 & 0.516 & 0.947 \\
    & L1 & 11.415 & 0.224 & 0.091 & 0.146 & -0.187 & 11.668 & 0.012 & 0.516 & 0.939 \\
    & L2 & 11.528 & 0.241 & 0.035 & 0.075 & -0.168 & 11.723 & 0.005 & 0.508 & 0.940 \\
    & L3 & 11.558 & 0.217 & 0.010 & 0.021 & -0.065 & 11.783 & 0.001 & 0.492 & 0.950 \\
    \hline
    $1.0 \leq z \leq 1.6$ & NB816 & 11.75 & 0.06 & 0.13 & 0.95 & - & 12.46 & 0.98 & 1 & 1.06\\
    \hline 
    $1.6 \leq z \leq 2.4$ & NB912 & 11.93 & 0.13 & 0.14 & 0.90 & - & 12.46 & 0.98 & 1 & 1.06\\
    \hline
    
    \end{tabular}
    \caption{The value of the HOD parameters in eq.~\ref{eq:HOD_parameter}. The first column shows in which redshift we used these HOD models, note it is not the same with the redshift range of the galaxy sample to fit these parameters. 
    The first four rows shows the fitting result in Table 4 of 
    ref.~\cite{Gao2022}. 
    And the last two rows shows the fitting results in Table 3 of 
    ref.~\cite{okumura2022a}.}
    \end{table*}

\section{The equivalence between kSZ measurement methods}
\label{appkSZMethodEquivalence}

The essence of measuring kSZ effect by cross-correlating a CMB and density survey is a bispecturm of temperature, large scale velocity (equivalent to density) and density, $\langle T\delta\delta \rangle$. 
 \cite{Smith2018a} has shown that the different methods of kSZ measurement are equivalent to each other formulaically. This section will briefly show the demonstration following the steps in \cite{Smith2018a}.
The starting is from the most general form of a three-point estimator
\begin{eqnarray}
\label{eq:kSZ_all_estimator}
    \hat{\xi} = \int \frac{d^3\mathbf k}{(2\pi)^3} \frac{d^3 \mathbf k'}{(2\pi)^3} \frac{d^2\mathbf l}{(2\pi)^2} W(\mathbf k, \mathbf k', \mathbf l) \left( \delta_g(\mathbf k) \delta_g(\mathbf k') T(\mathbf l) \right) 
    \nonumber \\ 
    {(2\pi)^3 \delta^3 (\mathbf k + \mathbf k' + \frac{\mathbf l}{\chi_*})}.
\end{eqnarray}
Next step is to determine the weighting function $W(\mathbf k, \mathbf k', \mathbf l)$, by minimizing the variance of $\hat \xi$. Requiring the estimator to be unbiased, the weight function would be
\begin{eqnarray}
    W(\mathbf k, \mathbf k', \mathbf l) = \frac{1}{2F_{BB}}\frac{-iB^*(k, k', l, k_r)}{P^{\text{tot}}_{gg}(k) P^{\text{tot}}_{gg}(k') C^{TT,\text{tot}}_l},
\end{eqnarray}
where $F_{BB}$ is the fisher matrix, $P^{\text{tot}}_{gg}(k)$ is the total power spectrum of the galaxies, $ C^{TT,\text{tot}}_l$ is the power spectrum of CMB including the instrument noise, and $B(k, k', l, k_r)$ is the bispectrum 
\begin{eqnarray}
    \langle \delta_g(\mathbf{k}) \delta_g(\mathbf{k}') T(\mathbf{l}) \rangle = B(\mathbf{k}, \mathbf{k}', \mathbf{l}) (2\pi)^3 \delta^3 \left( \mathbf{k} + \mathbf{k}' + \frac{\mathbf{l}}{\chi_*} \right).
\end{eqnarray}
It can be inferred that in the "squeezed" limit ($k$ is small, while $k'$ and $l$ are large), the tree-level kSZ bispectrum, which is assumed to be an accurate approximation of the kSZ bispecturm, is 
\begin{eqnarray}
    B\left(k_L, k_S, l, k_{L r}\right)=-\frac{K_* k_{L r}}{\chi_*^2} \frac{P_{g v}\left(k_L\right)}{k_L} P_{g e}\left(k_S\right)
\end{eqnarray}
Then the estimator becomes
\begin{eqnarray}\label{eq:kSZ_estimator_squeezed_limit}
    \hat{\mathcal{E}}&=&\frac{K_*}{\chi_*^2 F_{B B}} \int \frac{d^3 \mathbf{k}_L}{(2 \pi)^3} \frac{d^3 \mathbf{k}_S}{(2 \pi)^3} \frac{d^2 \mathbf{l}}{(2 \pi)^2} \frac{i k_{L r}}{k_L}\nonumber\\
        && \frac{P_{g v}\left(k_L\right) P_{g e}\left(k_S\right)}{P_{g g}^{\operatorname{tot}}\left(k_L\right) P_{g g}^{\operatorname{tot}}\left(k_S\right) C_l^{T T, \text { tot }}}
    \nonumber \\
    &&\left(\delta_g\left(\mathbf{k}_L\right) \delta_g\left(\mathbf{k}_S\right) T(\mathbf{l})\right)(2 \pi)^3 \delta^3\left(\mathbf{k}_L+\mathbf{k}_S+\frac{\mathbf{l}}{\chi_*}\right)
\end{eqnarray}
\begin{itemize}
    \item  \textbf{Prove stacking method is equivalent to the estimator}\\
    At the positions of galaxies, the velocity weighted stacking temperature is 
    \begin{eqnarray}
        \hat {\alpha} = \sum_i \hat{v}_i \Tilde{T_i},
    \end{eqnarray}
    where $\hat{v}_i$ is the reconstructed LOS velocity $\Tilde{T_i}$ is the filtered CMB temperature at position of galaxy $i$.
    The estimator (eq.~\ref{eq:kSZ_estimator_squeezed_limit}) can be rewrited with $\delta_g(\mathbf{k}_S) =\frac{1}{\Bar{n}_g} \sum_i e^{-i\mathbf{k}_S\cdot\mathbf{x}_i}$ as
    \begin{eqnarray}
        \hat{\mathcal{E}}= \frac{1}{\Bar{n}_g} \sum_i 
        \left(\int \frac{d^3 \mathbf{k}_L}{(2 \pi)^3}\frac{i k_{L_r}}{k_L} 
        \frac{P_{gv}(k_L)}{P_{gg}^{\rm{tot}}(k_L)}\delta_g(\mathbf{k}_L)  e^{-i\mathbf{k}_L\cdot\mathbf{x}_i}\right)
        \\ \nonumber \times
        \left(\int \frac{d^2 \mathbf{l}}{(2 \pi)^2} \frac{ P_{ge}(l/\chi_*)}{P_{gg}^{\text{tot}}(l/\chi_*) C_l^{TT, \text{tot}}} T(\mathbf{l})  e^{-i\mathbf{l}\cdot\mathbf{x}^{\perp}_i/\chi_*}\right)
    \end{eqnarray}
    The normalization ${K_*}/{\chi_*^2 F_{B B}}$ is removed for simplicity. Then the reconstructed velocity and the filtered temperature can be represented as
    \begin{eqnarray}
        \hat v_i = \int \frac{d^3 \mathbf{k}_L}{(2 \pi)^3}\frac{i k_{L_r}}{k_L} \frac{P_{gv}(k_L)}{P_{gg}^{\rm{tot}}(k_L)}\delta_g(\mathbf{k}_L)  e^{-i\mathbf{k}_L\cdot\mathbf{x}_i}\\
        \Tilde{T_i} = \int \frac{d^2 \mathbf{l}}{(2 \pi)^2} \frac{ P_{ge}(l/\chi_*)}{P_{gg}^{\text{tot}}(l/\chi_*) C_l^{TT, \text{tot}}} T(\mathbf{l})  e^{-i\mathbf{l}\cdot\mathbf{x}^{\perp}_i/\chi_*}.
    \end{eqnarray}
    In addition, the above equations also point out the optimal filter for the reconstructed velocity and the filtered temperature.

    \item \textbf{Prove tomography method is equivalent to the estimator}\\
    The kSZ tomography template is the product of the reconstructed velocity and density along the line of sight
    \begin{eqnarray}
        \hat \Theta = \int^{z_2}_{z_1} dz \hat{v} \hat{\delta},
    \end{eqnarray}
    where
    \begin{eqnarray}
        \hat{v} &=& \frac{i k_{L_r}}{k_L} \frac{P_{gv}(k_L)}{P_{gg}^{\rm{tot}}(k_L)}\delta_g(\mathbf{k}_L) \\
        \hat{\delta} &=& \frac{P_{ge}(k_S)}{P_{gg}^{\rm{tot}}(k_S)}\delta_g(\mathbf{k}_S)
    \end{eqnarray}
    Here,
    \begin{eqnarray}
        W_1 =  \frac{P_{gv}(k_L)}{P_{gg}^{\rm{tot}}(k_L)}\\
        W_2 =  \frac{P_{ge}(k_S)}{P_{gg}^{\rm{tot}}(k_S)}
    \end{eqnarray}
    are the optimal filters to reconstruct $\hat v$ and $\hat \delta$
    The template $\hat \Theta $ in Fourier space is 
    \begin{eqnarray}
        \hat T(\mathbf{l}) = \int \frac{d^3 \mathbf{k}_L}{(2 \pi)^3} \frac{d^3 \mathbf{k}_S}{(2 \pi)^3} \frac{i k_{L r}}{k_L} \frac{P_{g v}\left(k_L\right)}{P_{g g}^{\operatorname{tot}}\left(k_L\right)} \frac{P_{g e}\left(k_S\right)}{P_{gg}^{\operatorname{tot}}\left(k_S\right)}
        \nonumber \\
        \left(\delta_g\left(\mathbf{k}_L\right) \delta_g\left(\mathbf{k}_S\right)\right)(2 \pi)^3 \delta^3\left(\mathbf{k}_L+\mathbf{k}_S-\frac{\mathbf{l}}{\chi_*}\right)
    \end{eqnarray}
    And the estimator becomes to 
    \begin{eqnarray}
        \hat{\mathcal{E}} =  \int \frac{d^2\mathbf{l}}{(2 \pi)^2}\frac{1}{C_l^{TT, \text{tot}}} \left( T(\mathbf{l})T(-\mathbf{l})\right),
    \end{eqnarray}
    where $T(\mathbf{l})T(-\mathbf{l})$ is the correlation between the template $\hat \Theta$ and the true kSZ signal $\Theta$.
\end{itemize}

\begin{figure*}
    \centering
	\includegraphics[width=0.7\textwidth]{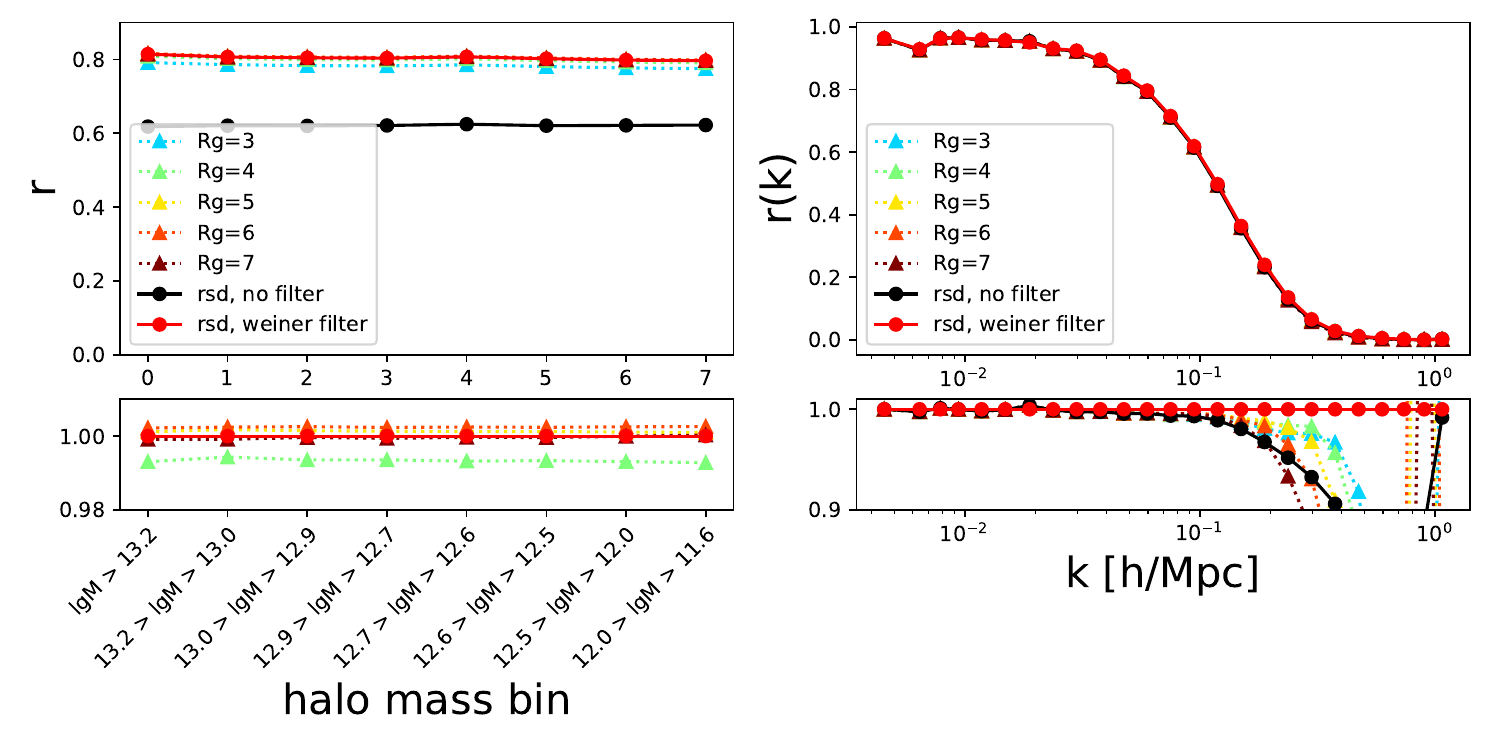}
    \includegraphics[width=0.7\textwidth]{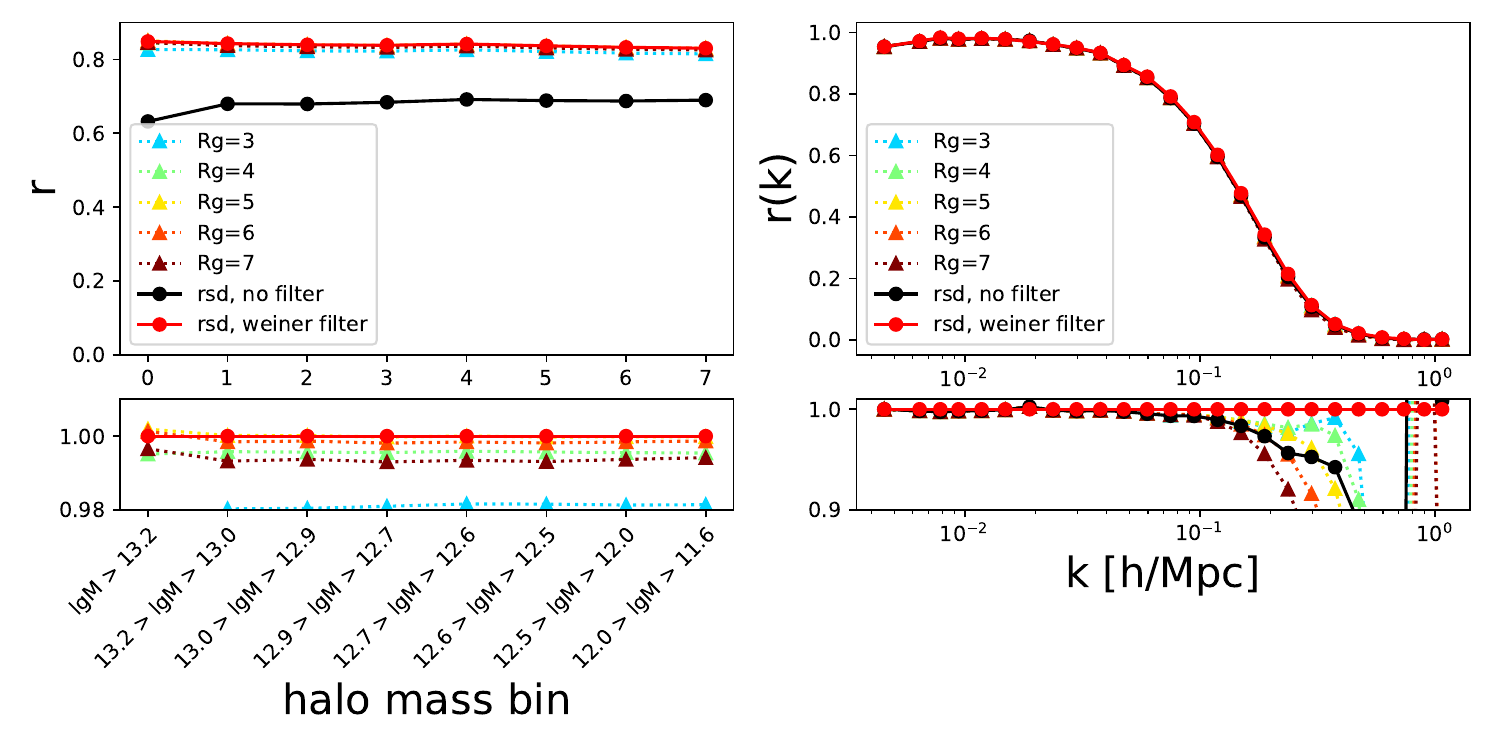}
    \includegraphics[width=0.7\textwidth]{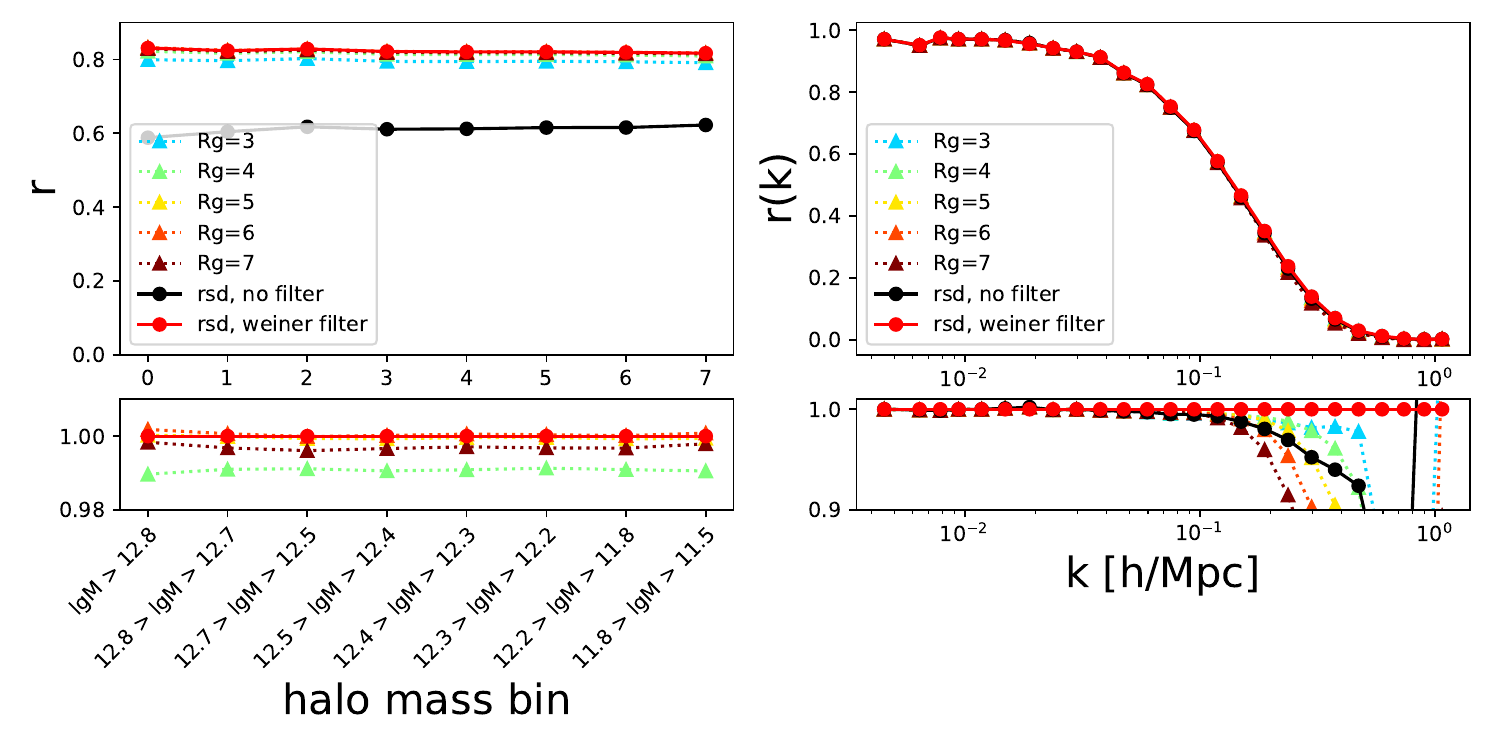}
	\caption{The correlation coefficient of different galaxy group mass bins (left) and the correlation function (right) of the true and reconstructed velocity fields in three redshift bins $0.6 < z < 1.0$, $1.0 < z < 1.6$ and $1.6 < z < 2.4$ (from top to bottom). The black lines show the case that reconstructing velocity field directly from $\delta_g$ without applying any filters. The red lines show the case of applying the Wiener filter using the main text. The dashes lines show the case of applying the Gaussian filter with different radius $R_g$. \label{fig:rk_filter}}
\end{figure*}
\section{Test the velocity reconstruction method}
\label{app:test_vel_rec}
In this section, our research delves into the impact of applying different filters to the density field on velocity reconstruction performance. In the previous literatures,  Gaussian filters in Fourier space, represented as $W_g(k) = \exp(-k^2R_s^2)$, have been utilized to mitigate shot noise in galaxy density fields, with $R_s$ being a pivotal, adjustable parameter for optimizing velocity reconstruction.
In figure \ref{fig:rk_filter}, we present a comparative analysis of the correlation coefficients across various mass bins of galaxy groups and the correlation function between true and reconstructed velocities. Our findings suggest that, in the stacking method, the performance of the optimal Gaussian filter is comparable to the Wiener filter discussed in the main text. However, the Wiener filter demonstrates a noticeable advantage in the tomography method at smaller scales $k>0.1 [h/{\rm Mpc}]$, highlighting the Wiener filter's efficacy in enhancing velocity reconstruction at these scales.

%This is the most common positions for %acknowledgments. A macro is
%available to maintain the same layout %and spelling of the heading.

%\paragraph{Note added.} This is also a good position for notes added after the paper has been written.

% Bibliography

%% [A] Recommended: using JHEP.bst file
%% \bibliographystyle{JHEP}
%% \bibliography{biblio.bib}

%% or
%% [B] Manual formatting (see below)
%% (i) We suggest to always provide author, title and journal data or doi:
%% in short all the informations that clearly identify a document.
%% (ii) please avoid comments such as "For a review'', "For some examples",
%% "and references therein" or move them in the text. In general, please leave only references in the bibliography and move all
%% accessory text in footnotes.
%% (iii) Also, please have only one work for each \bibitem.
\bibliographystyle{JHEP}
\bibliography{kSZ_forecast.bib}

\end{document}